\documentclass[12pt]{iopart}
                                                                   
\usepackage{graphicx}
\usepackage{placeins}
\usepackage{subfigure}
\usepackage{wasysym}
\usepackage{textcomp}
\usepackage[utf8]{inputenc}
\usepackage{epsfig,graphics}
\usepackage{amssymb}
\usepackage{subfigure}
\newcommand{\notes}[1]{}

\begin{document}

\title{Detrended Fluctuation Analysis of Systolic Blood Pressure Control Loop}

\author{C.E.C. Galhardo$^1$, T.J.P.Penna$^1$, M. Argollo de Menezes$^1$
  and P.P.S. Soares$^2$ } \address{$^1$ Instituto de F\'{\i}sica,
  Universidade Federal Fluminense, Av. Litoranea, s/n, 24210-340,
  Niteroi, RJ ,Brazil } \address{$^2$ Instituto Biom\'{e}dico,
  Universidade Federal Fluminense, R. Prof. Hernani Melo n. 101,
  24210-130, Niteroi, RJ, Brazil } \date{\today}

\begin{abstract}
  We use detrended fluctuation analysis (DFA) to study the dynamics of
  blood pressure oscillations and its feedback control in rats by
  analyzing systolic pressure time series before and after a surgical
  procedure that interrupts its control loop. We found, for each
  situation, a crossover between two scaling regions characterized by
  exponents that reflect the nature of the feedback control and its
  range of operation. In addition, we found evidences of adaptation in
  the dynamics of blood pressure regulation a few days after surgical
  disruption of its main feedback circuit. Based on the paradigm of
  antagonistic, bipartite (vagal and sympathetic) action of the
  central nerve system, we propose a simple model for pressure
  homeostasis as the balance between two nonlinear opposing forces,
  successfully reproducing the crossover observed in the DFA of actual
  pressure signals.
\end{abstract}

\submitto{\NJP}

\maketitle

\section{Introduction}
Negative feedback loops are ubiquitous in living systems, with
important examples like the lac-operon in gene regulation
\cite{JMolBiol:3:318}, which inhibits lactose consumption in the
presence of glucose, and serve as efficient ways of maintaining
stability and suppressing fluctuations in noisy environments
\cite{Nat:252:546,Nat:405:590,Nat:427:415,MolSysBio:2:41,AJP_RICP:291:R1638}
\footnote{ Negative feedback loops also appear in electronic circuits
  as a tool for the stabilization of laser beams (see
  \cite{OptLett:5:191})}.  On a much larger physical scale, the
autonomous nerve system is able to sustain (without external
supervision) basic life signals like temperature, water and
metabolite concentrations at safe levels by the action of a pair of
nerve branches, called sympathetic and parasympathetic (or vagal).
These nerve branches have cooperative and ``antagonistic'' roles in
our body: while the sympathetic prepares our body for
``flight-or-fight'' situations (increasing heart rate, dilating pupils
and cancelling digestive functions, for instance), the vagal, or
parasympathetic, decreases heart rate, constricts pupils and stimulate
salivary glands. The balance between these ``forces'', which keeps
living systems operating close to optimal levels, is called {\it
  homeostasis} \cite{book_phys,book_phys1}. Alterations of a given control
mechanism can perturb such balance and lead to pathological conditions
such as Diabetes Mellitus, which results from a malfunctional insulin
metabolism \cite{AJP_RICP:291:R1638}.

\begin{figure}[!hbt]
\centering
\includegraphics[scale=0.3]{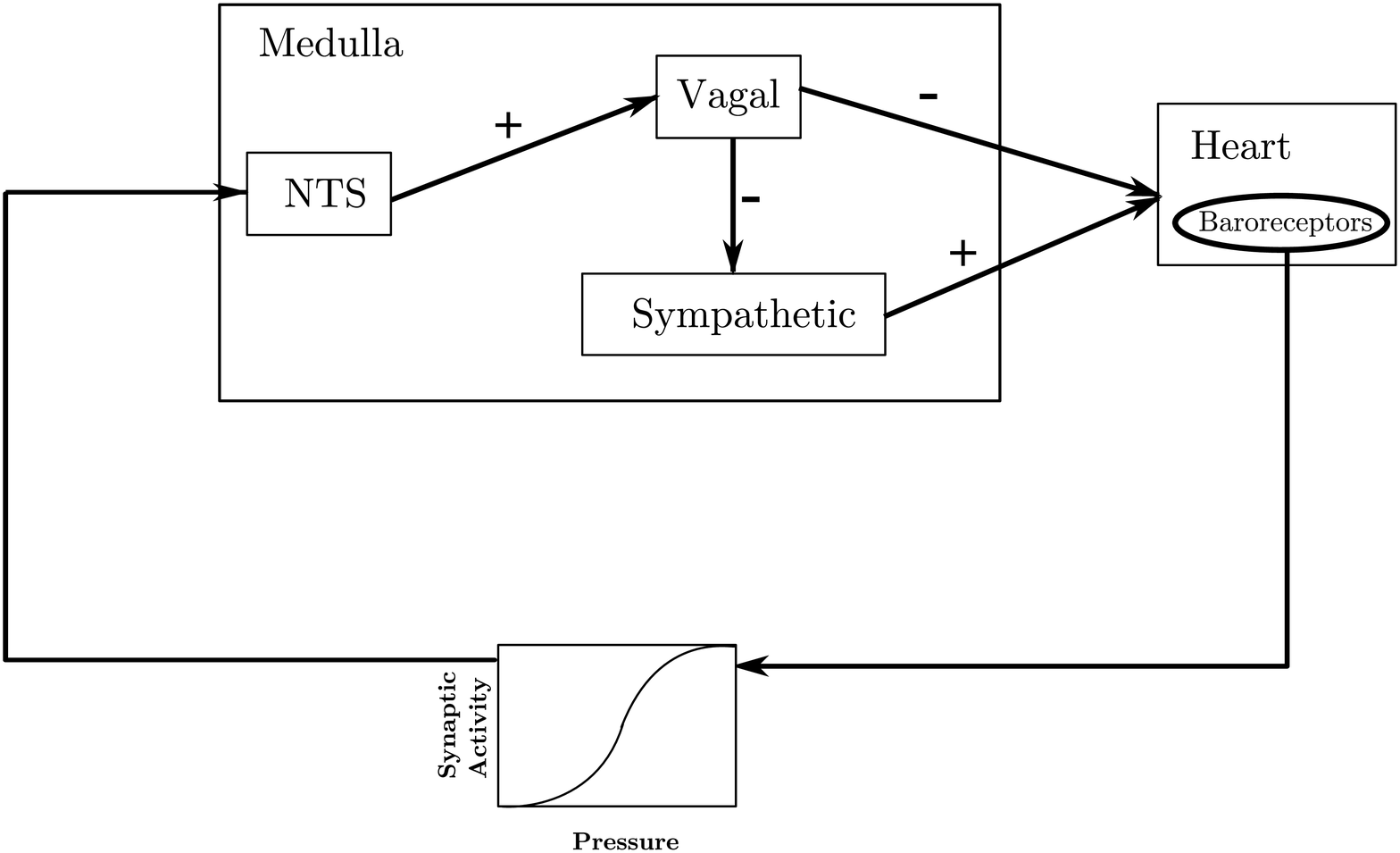}
\caption{Schematics of the negative feedback loop for pressure
  control, or baroreflex. Stimulus from afferent neurons excite the
  vagal branch of the autonomous nerve system, which in turn slows
  down heart rate. At the same time, the sympathetic branch, which
  acts to increase heart rate, is inhibited by the vagal branch. As a
  result, a surge in blood pressure tends to stimulate the vagal
  branch and inhibit the sympathetic branch, decreasing heart rate and,
  consequently, decreasing blood pressure.}
\label{fig:baro}
\end{figure}

A major feature of the autonomous nerve system is that stimulation of
the vagal branch results in a inhibition of the sympathetic branch,
which acts continuously on organs and veins at an approximately steady
level when not inhibited.  These nerve branches are controlled at the
Nucleus Tractus Solitarius (NTS) of the medulla by integration of
neural information coming from afferent neural fibers, which carry
information from sensory neurons spread around the body. Among other
sensory information carried by those fibers, one of utmost importance
regards arterial blood pressure: through these afferent nerve fibers,
stretch-sensitive mechanoreceptors spread around veins and arteries of
the heart return to the NTS (in a timescale of few seconds)
information about the current status of pressure (and its
variation). \footnote[1]{There are also baroreceptors at the kidneys,
  which change body fluid volume at the timescale of hours or days
  \cite{Sci:252:1813}.  Those are responsible for very low frequency
  fluctuations and will not be analyzed here.} The NTS, in turn,
excite (when pressure is high) or inhibit (when pressure is low) the
vagal branch, closing the circuit for what can be regarded as a
self-inhibitory feedback loop called {\it baroreflex}
\cite{book_phys,book_phys1,RegIntCompPhys:288:R815} (See figure \ref{fig:baro}).

\begin{figure}[!hbt]
\centering
\includegraphics[scale=0.5]{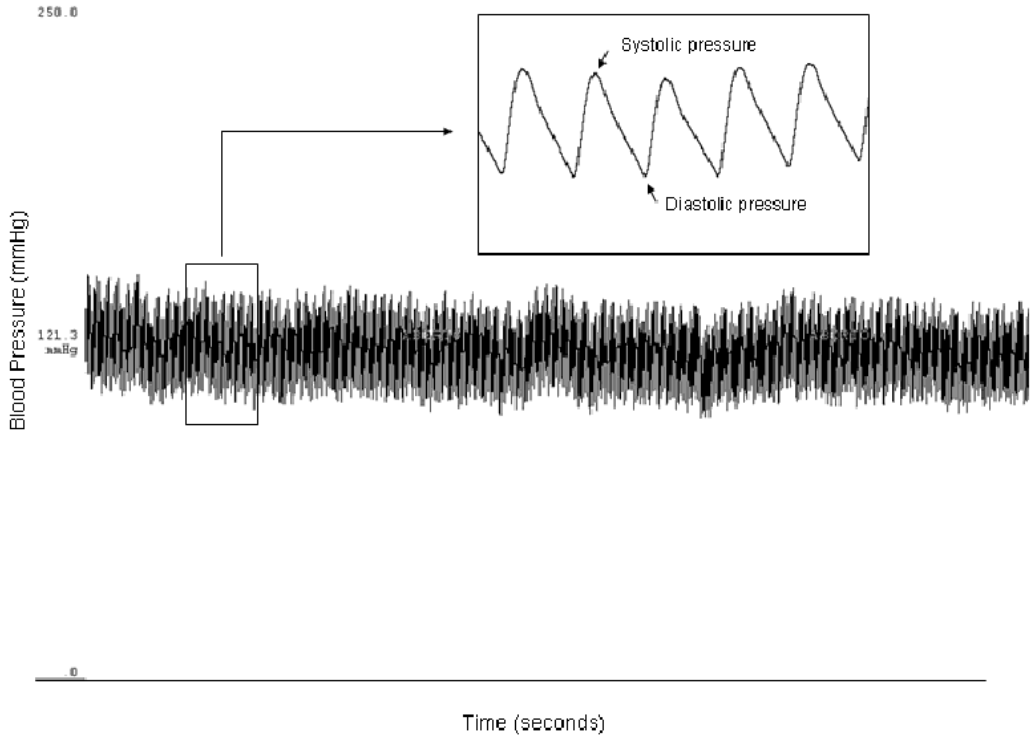}
\caption{Time evolution of arterial blood pressure (ABP). The local
  maxima are called systolic blood pressure, the minima are the diastolic
  pressure and the time interval between two neighboring ABP minima is
  the instantaneous inter-beat heart rate. In this work we focus on
  the arterial systolic blood pressure and its variation in time.}
\label{fig:bloodwave}
\end{figure}

As a result of this balance the body, although continuously perturbed
by external factors, is able to keep homeostasis, a stationary state
where, among other things, arterial pressure, temperature, water and
metabolite concentrations are kept at optimal levels
\cite{book_phys,book_phys1}.  One can think of homeostasis as a
locally optimal state sustained by feedback loops in a noisy
environment. The reasonably controlled flow of nutrients throughout
veins and arteries is achieved with the aid of the blood system and
the heart, whose pumping action is monitored and controlled by the
autonomous nerve system.  Arterial blood pressure (ABP) is one of the
vital signals that can be continuously monitored, which carries a
large amount of information about the mechanisms responsible for
homeostasis and the different timescales for their responses
\cite{JApplPhysio:101:676,Hyper:51:811}. Given a continuous set of
recordings of ABP, $\{p(t)\}$, over a given period of time, one
defines the $n$-th diastolic blood pressure as the $n$-th local
minimum $p_n$, the systolic blood pressure as the $n$-th local maximum
$p_n$ and the time interval between two neighboring ABP minima,
$b_n=t(p_{n+1})-t(p_{n})$, as the instantaneous inter-beat heart rate
(IR), as depicted in figure \ref{fig:bloodwave}.

These quantities have long been characterized by spectral methods
\cite{AutNeurSci:113:24,Sci:213:220a}, where peaks in the power
spectrum $S_{\omega}=\left |\frac{1}{2\pi}\sum b_n e^{i\omega
  n}\right|^2$ reveal the timescales for the response of different
control mechanisms
\cite{JHyper:2:S383,Hyper:12:600,JHyper:7:S32,Circ:86:1443,AJP_HCP:276:H1987}.
Nevertheless, in order to assess the long-range correlations
\cite{PhysRevLett:70:1343} emerging from these feedback control
systems, or to characterize disruptive and abnormal states, one must
recur to methods which account for the strong non-stationarity of
those signals \cite{PhysioMeas:23:R1}, such as detrended fluctuation
analysis (DFA) \cite{PhysRevE:49:1685,PhysRevE:65:041107,Chaos:5:82,
  PhysioMeas:25:763,PhysRevE:66:2902,PhysRevLett:85:3736,PhyA:384:429}. In
this work we analyze the dynamics of baroreflex, the negative feedback
loop providing a rapid and powerful reflex control of blood pressure,
which is by far the most studied cardiovascular reflex in
physiological and clinical settings. For such purpose we apply DFA to
experimental time series consisting of continuous arterial systolic
blood pressure measurements.

We report results of experiments on rats with surgical disruption of
the nerve fibers connecting the baroreceptors to the medulla, a
procedure called sinoaortic denervation (SAD)
\cite{RegIntCompPhys:266:R1705}, and find that other mechanisms might
be responsible for arterial blood pressure control, although at
different time scales, possibly due to synaptic plasticity at the NTS
\cite{AJP_RICP:289:R1416,JApplPhysio:101:322,RespPhysio:122:83}.
Following this recovery, average blood pressure is kept at almost the
same levels as before denervation, a determinant condition for the
kidneys to work properly \cite{book_phys,book_phys1}.  We apply
detrended fluctuation analysis to our experimental time series and
find that fluctuations in systolic blood pressure cross over from
non-stationary to stationary, long-range correlated at a
characteristic time scale $\tau$.  Surgical denervation of
baroreceptors significantly changes the correlation patterns of
pressure signals but, after $20$ days, correlation patterns typical of
non-operated rats are recovered, only with larger crossover times
$\tau^{\prime}>\tau$.  This suggests that the control loop is
reestablished, possibly due to adaptation to sensory information
coming from other less effective receptors.

To model such feedback control loop we develop a model of a random
walker forced by two opposing nonlinear (sigmoidal) forces,
representing the sympathetic action and its inhibition by the vagal
(parasympathetic) branch. We find the same crossover from
non-stationary to stationary, long-range correlated noise observed in
actual pressure measurements. Moreover, by changing the difference
between the sensitivity of each branch, we find the same shift in the
crossover time scale, as observed in rats $20$ days after surgery,
when adaptation occurs and homeostasis is recovered.


\section{Experiments and Measurements}

Adult male Wistar rats were maintained on a $12$-hour light/dark
lighting schedule at $23^o$C, food and water \textit{ad libitum}. All
procedures were performed according to \cite{AnimalCare}. The animals
were divided in three groups: control rats (ctr, $N=11$ rats), acute
sinoaortic denervated rats (1d, $N=5$ rats), i.e, animals surgically
denervated one day before measurements, and chronic sinoaortic
denervated rats (20d, $N=8$ rats), animals surgically denervated $20$
days before measurements. SAD was performed using the methods
described by Krieger et al \cite{CircRes:15:511}, and basically
consists of full disruption of the nerve fibers connecting the
baroreceptors spread in veins and arteries of the heart to the
medulla.  Blood pressure was recorded from the left femoral artery for
$90$ minutes in conscious rats.  Before the analog to digital
conversion, blood pressure was low-pass filtered (fc= $50$ Hz) for
high-frequency noise removal, and recorded with a $2$kHz sampling
frequency.  Systolic (maximum) and diastolic (minimum) values were
detected after parabolic interpolation and signal artifacts were
visually identified and removed. Pulse intervals were measured in
milliseconds (ms), considering intervals between consecutive diastole
and the heart rate was calculated as the inverse of pulse interval and
measured in beats per minute (bpm) (A more detailed account of this
experiment can be found in \cite{JCardPhar:47:331}).  Since the
measurements were made in awake, conscious unrestrained rats, some
distortions in the blood pressure signal might arise due to their
movements.  To reduce this problem we discard series that show any
kind of discontinuities or jumps. After this selection we keep six
time series for the control group, five time series for the chronic
denervated group and four time series for the acute denervated
group. Each time series consists of $10^4$ data points, equivalent to
$30$ minutes of continuous measurements.

\begin{figure}[ht]
\centering
\subfigure[]{
\includegraphics[scale=0.3,angle = 270]{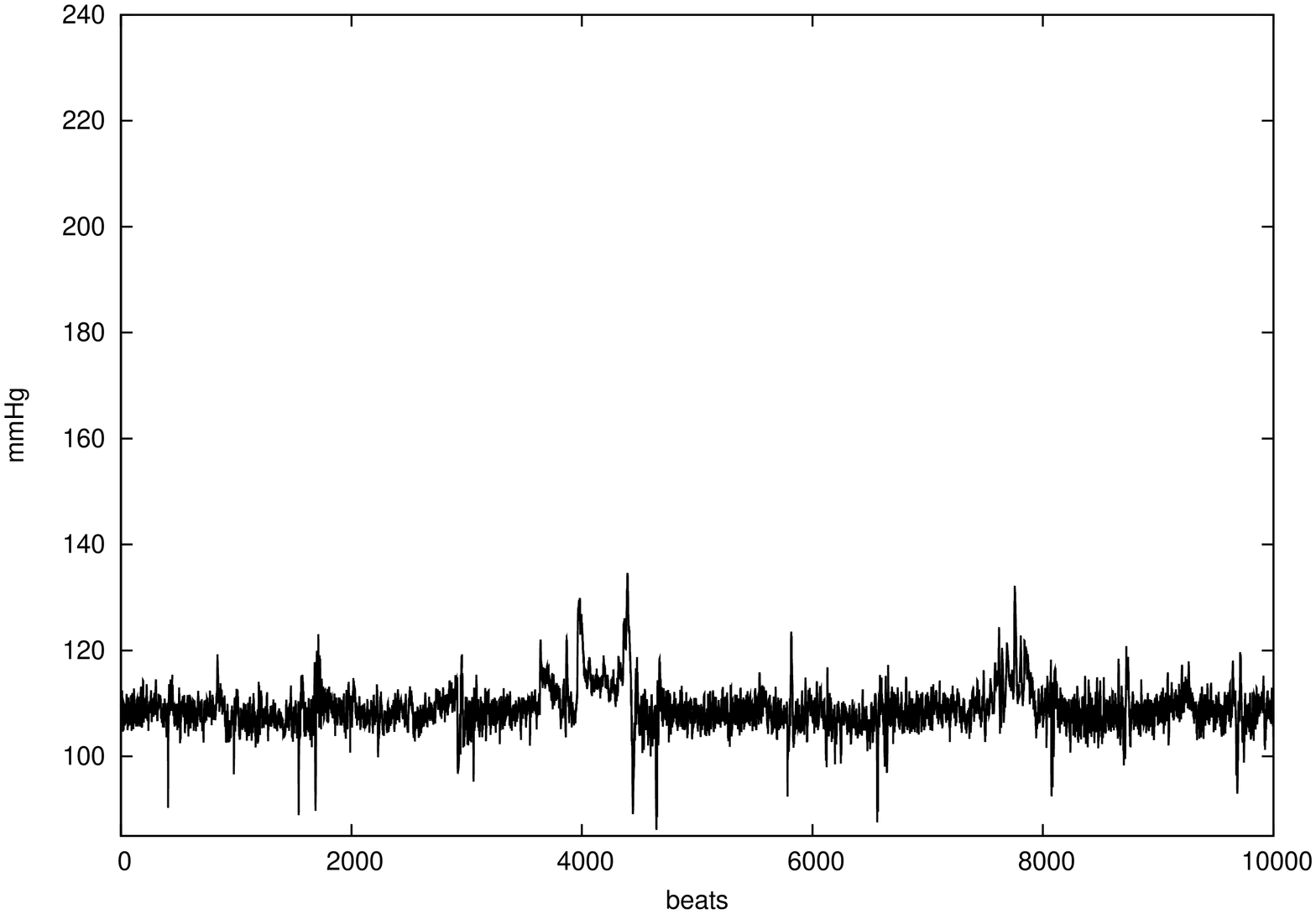}
\label{fig:seriesctr}
}
\subfigure[]{
\includegraphics[scale=0.3,angle = 270]{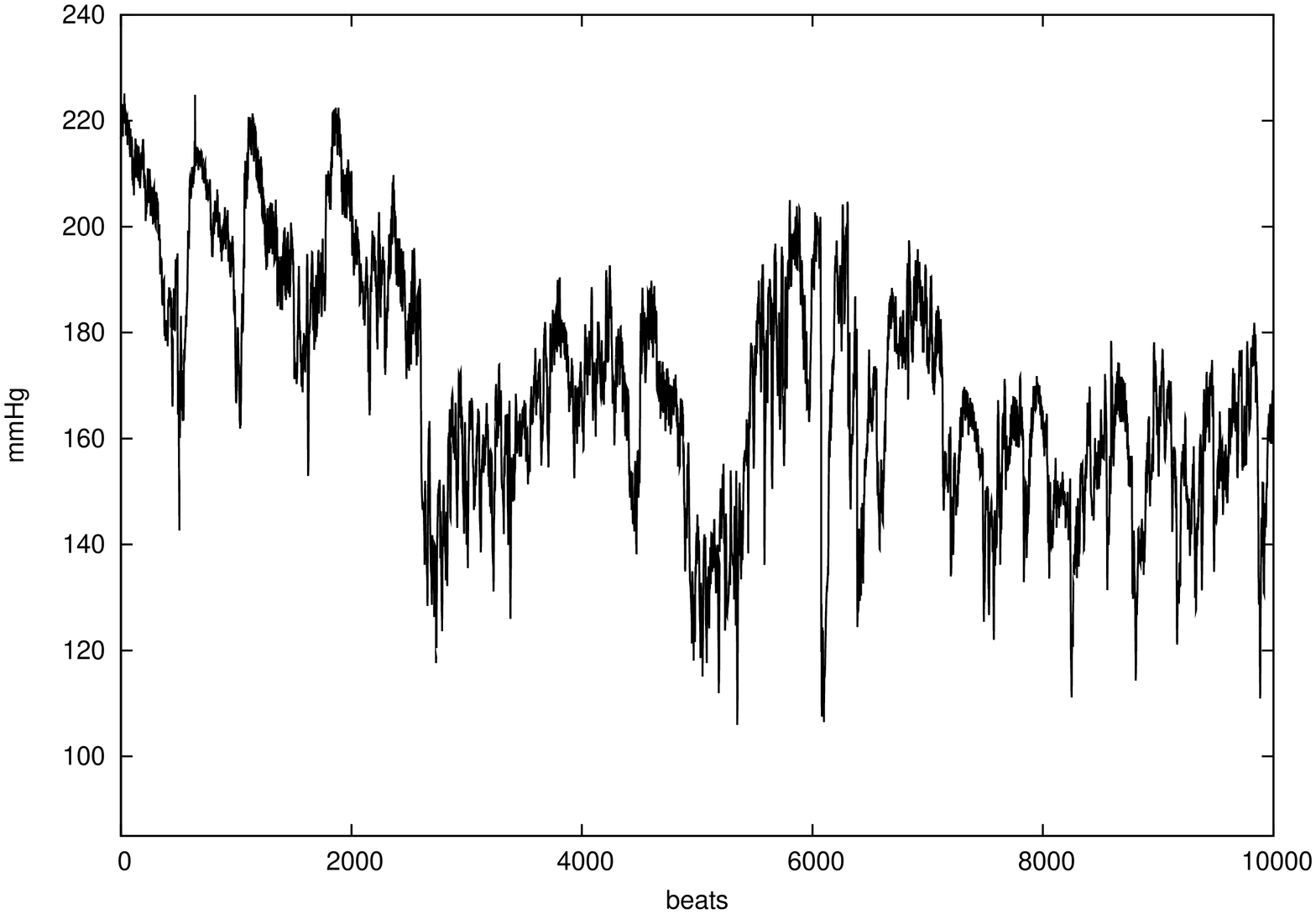}
\label{fig:series1d}
}
\subfigure[]{
\includegraphics[scale=0.3,angle = 270]{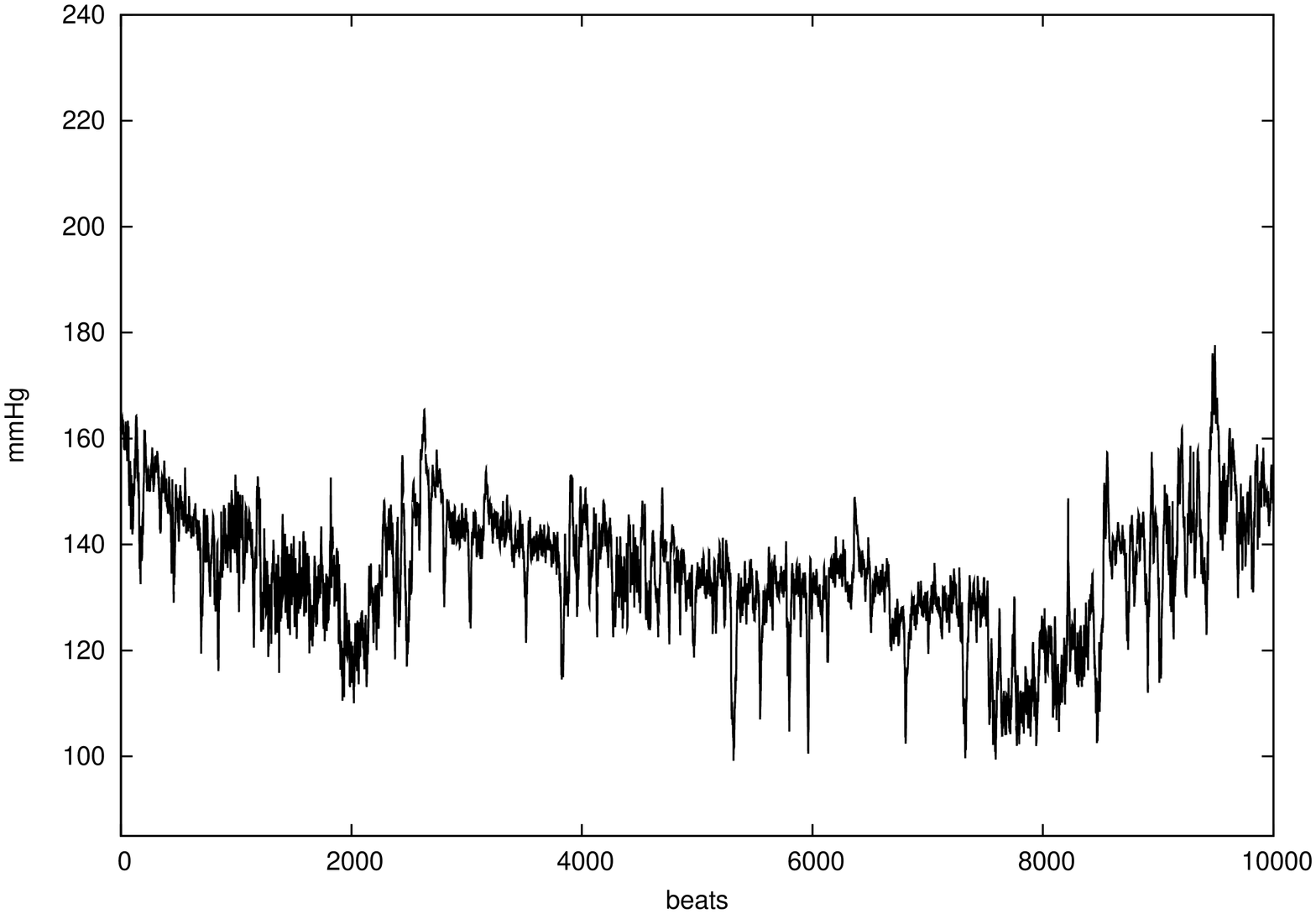}
\label{fig:series20d}
}
\caption{{\bf \subref{fig:seriesctr}} Fluctuations of arterial
  systolic blood pressure from a rat in the control group. Blood
  pressure oscillates about safe, steady levels. {\bf
    \subref{fig:series1d}} One day after disrupting the pressure
  control loop with a surgical procedure, pressure fluctuates in a
  non-stationary fashion, reaching dangerously high values.  {\bf
    \subref{fig:series20d}} As a result of physiological adaptation,
  20 days after surgical denervation of baroreceptors average blood
  pressure returns to safe levels and fluctuations are again
  stationary.}
\label{fig:seriesExamples}
\end{figure}

In figure \ref{fig:seriesExamples} we depict the series of systolic
blood pressure values for the three groups: while pressure in
non-operated rats fluctuates in a stationary fashion about $116.55 \pm
10.15$ mm Hg (Figure \ref{fig:seriesExamples}a), it is non-stationary
in rats with disrupted baroreflex (Figure \ref{fig:seriesExamples}b),
fluctuating about a much higher average value of $178.31 \pm 31.15$ mm
Hg.  After a period of $20$ days, average blood pressure falls back to
safe levels, $129.95 \pm 9.32$ mm Hg, and fluctuations are again
stationary (Figure \ref{fig:seriesExamples}c), indicating that
baroreflex is recovered. In order to understand the underlying
principles behind blood pressure regulation and the sources of
fluctuations in blood pressure levels we give, in the next section, a
precise, quantitative meaning to such fluctuations with detrended
fluctuations analysis (DFA).

\section{Fluctuation Analysis and Computer Modelling}

We used detrended fluctuation analysis (DFA)
\cite{Chaos:5:82,PhysRevE:49:1685} to characterize long term
correlations in arterial systolic blood pressure. This method has been
successfully applied to analyze diverse non-stationary physiological
signals
\cite{PhyA:274:99,AJPhysioRegIntCom:293:R1923,Chaos:5:82,PhysioMeas:25:763,PhysRevE:66:2902,PhysRevLett:85:3736}
and we briefly describe it in the following: Let $\{P(t)\}$ be the
systolic blood pressure time series and $P_{ave}$ its time
average. Define the integrated time series $\{y(t)\}$ with
\begin{equation}
y(t)=\sum_{k=1}^t (P(k)-P_{ave})
\end{equation}

\noindent Divide the integrated series in boxes of equal sizes $n$
and, for each box, calculate the detrended profile subtracting from
the original signal a $l$-degree polynomial least-squares fit,
$y^l_n(t)$ (In the following DFA$-l$ will stand for detrended
fluctuation analysis with $l$-degree polynomials
\cite{PhyA:387:5080}). At each box of size $n$, calculate the
fluctuation
\begin{equation}
F(n)=\sqrt{\frac{1}{N}\sum_{t=1}^N \left(y(t)-y^l_n(t)\right)^2}.
\end{equation}
A power-law relation $F(n)\sim n^{\alpha}$ implies different
correlation patterns for different values of $\alpha$: When
$0<\alpha<1/2$ the signal is stationary and long-range
anti-correlated, with $\alpha = 1/2$ for a white noise (and
$\alpha=3/2$ for its integral, the Brownian motion), $\alpha>1/2$ for
long-range correlated signals, while the paradigmatic $1/f$ noise
corresponds to $\alpha=1$. This value of $\alpha$ also marks the
borderline between stationary and non-stationary behavior: For
$\alpha\ge 1$ one has non-stationary signals, with sub-diffusive
($\alpha<3/2$), diffusive ($\alpha=3/2$) or super diffusive
($\alpha>3/2$) behavior.

\begin{figure}[!htb]
\centering
\subfigure[]{
\includegraphics[scale=0.36]{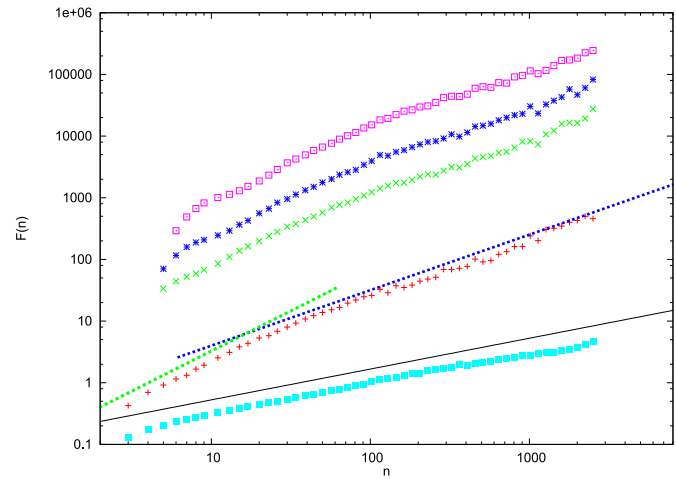}
\label{fig:dfaCtrTip}
}
\subfigure[]{
\includegraphics[scale=0.23]{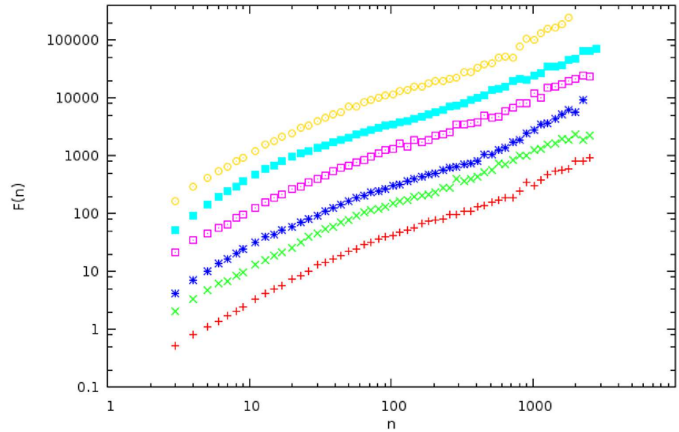}
\label{fig:dfaCtrAll}
}
\caption{ {\bf\subref{fig:dfaCtrTip}} Detrended fluctuation analysis
  of systolic blood pressure time series for a typical rat in the
  control group. There is a crossover from non-stationary to
  stationary, long-range correlated behavior at $n \approx 35$: For
  short time scales we have $\alpha \approx 1.18$ and for large time
  scales $ \alpha \approx 0.93$.  We apply DFA-1 (red crosses), DFA-2
  (green times), DFA-3 (blue stars) and DFA-4 (pink empty squares) to
  the series and find that the crossover always exists, although at
  different scales.  We also applied DFA-1 to shuffled data (bottom
  curve), for which $\alpha \approx 0.5$ as for a white
  noise. {\bf\subref{fig:dfaCtrAll}} DFA-1 for all rats in the control
  group. In both figures curves are shifted vertically for better
  visibility. The curves $y=Ax^{\alpha}$ with $\alpha=0.5$ (full black
  line), $0.9$ (dashed blue line) and $1.3$ (dashed green  line) are
  plotted as guides to the eye.}
\label{fig:dfaCtrSist}
\end{figure}

Results for a typical time series from the control group are depicted
in figure \ref{fig:dfaCtrSist}a.  With DFA-1 we obtain a crossover
from $\alpha = 1.18$ to $\alpha = 0.93$ at $n \approx 35$.  To check
that the crossover is not an artifact of a specific polynomial fit or
non-stationarities \cite{PhysRevE:64:011114,PhysRevE:65:041107,PhyA:295:441},
we also employed DFA-2, DFA-3 and DFA-4 on the time series. For all
orders $l$ there is a crossover, although at slightly shifted time
scales. We also show surrogate data, where data points are randomly
shuffled, and applied DFA-1 to it (Figure \ref{fig:dfaCtrSist}a,
bottom curve) to find that fluctuations scale with $\alpha\approx
0.5$, as in a typical white noise. We depict in figure
\ref{fig:dfaCtrSist}b results for all rats in the control group,
evidencing the same behavior in all curves.

\begin{figure}[!htb]
\centering
\subfigure[]{
\includegraphics[scale=0.38,angle=270]{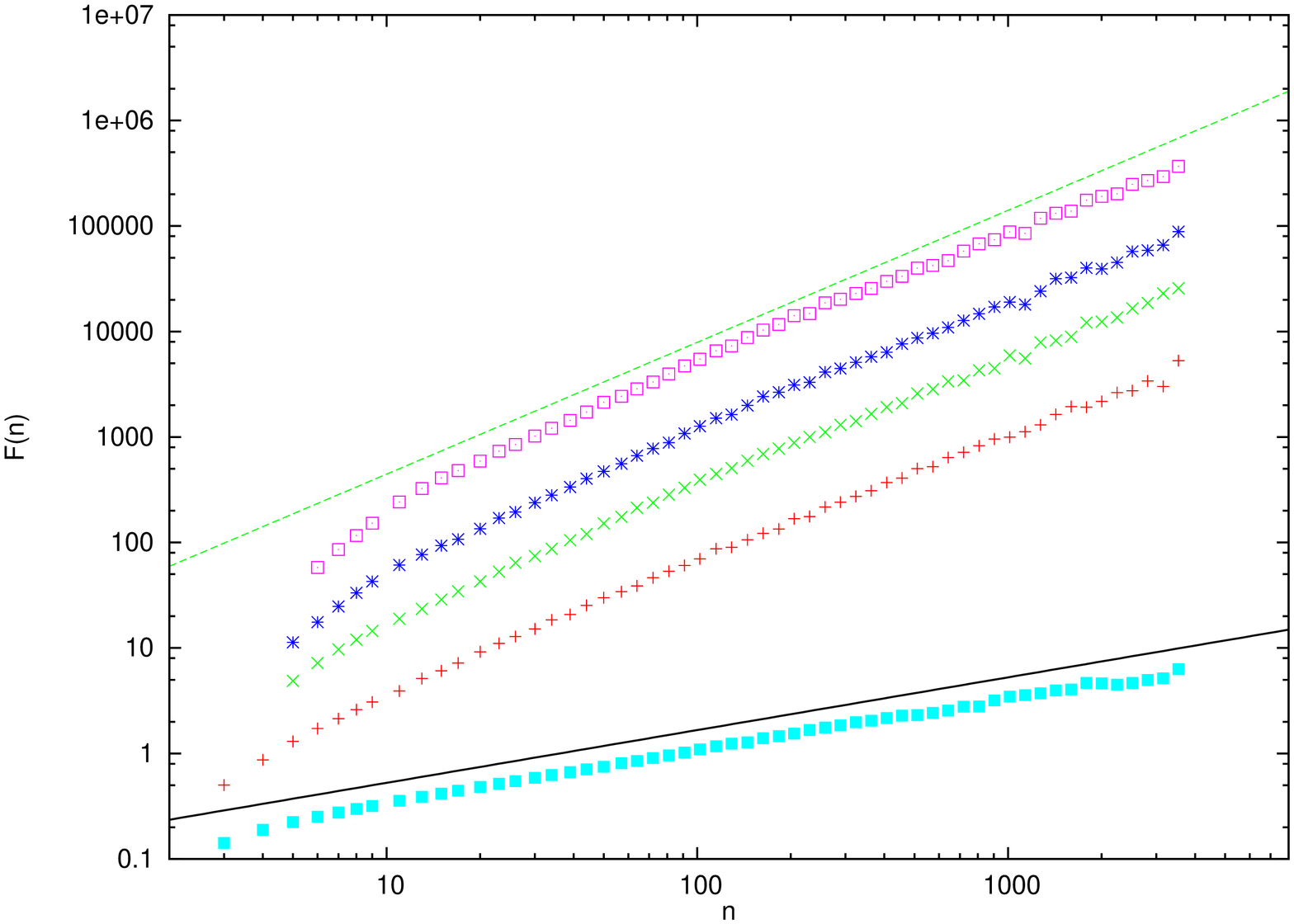}
\label{fig:dfa1dTip}
}
\subfigure[]{
\includegraphics[scale=0.45]{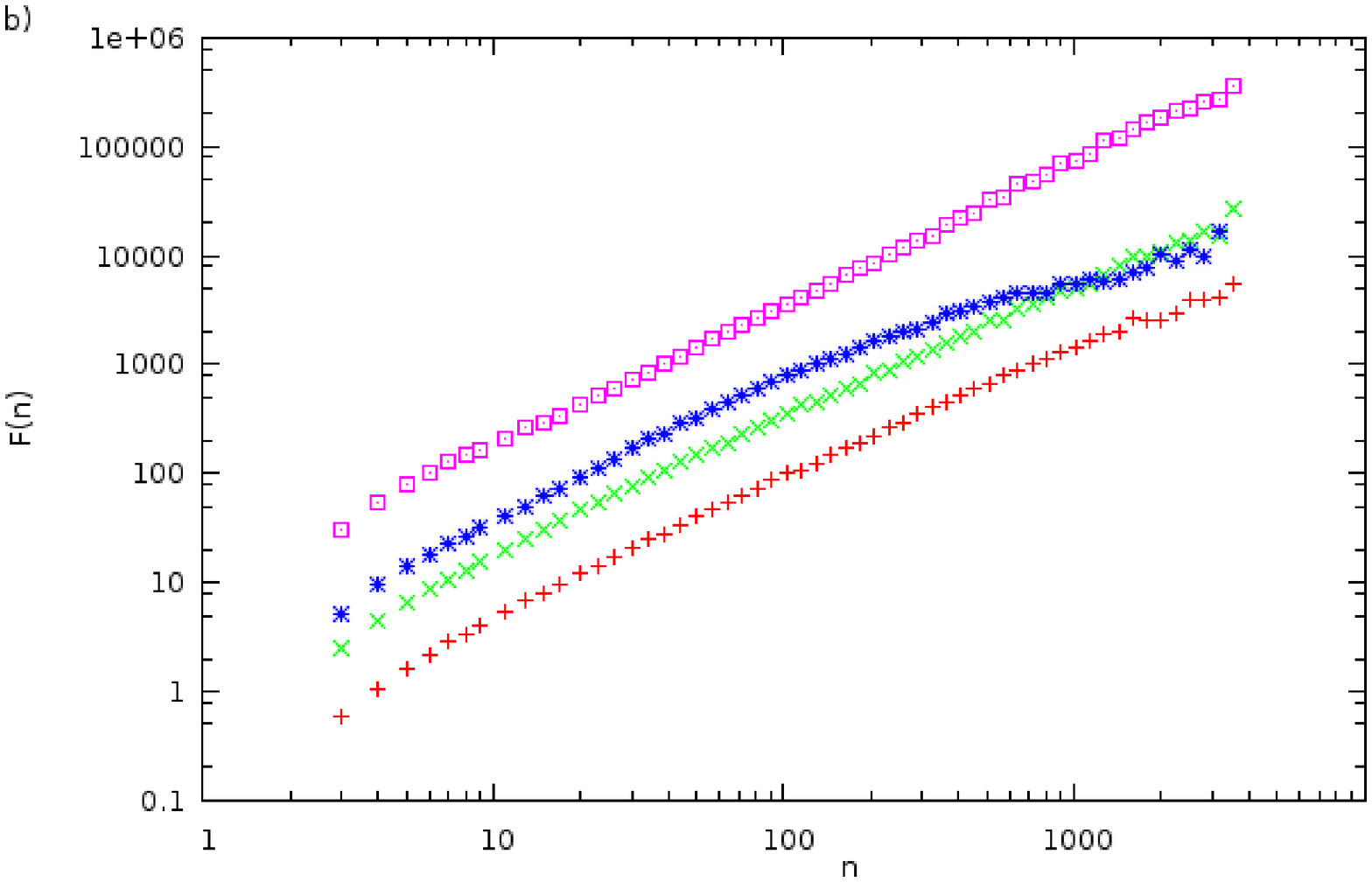}
\label{fig:dfa1dAll}
}

\caption{ {\bf\subref{fig:dfa1dTip}} Detrended fluctuation analysis of
  systolic blood pressure time series for a typical rat in the acute
  group. One day past surgical denervation of baroreceptors,
  fluctuations in blood pressure are non-stationary at all time scales
  and there is no crossover in the $F(n)$ curve. We apply DFA-1 (red
  crosses), DFA-2 (green times), DFA-3 (blue stars) and DFA-4 (pink
  empty squares) and find $\alpha \approx 1.25$, indicating a
  disruption of short-term homeostatic control of blood pressure, or
  baroreflex. We also applied DFA-1 to shuffled data (bottom
  curve,full squares), for which $\alpha \approx 0.5$ as for white
  noise. {\bf\subref{fig:dfa1dAll}} DFA-1 for all rats in the acute
  group. In both figures curves are shifted vertically for better
  visibility. The curves $y=Ax^{\alpha}$ with $\alpha=0.5$ (full black
  line) and $1.25$ (dashed green line) are plotted as guides to the
  eye.}
\label{fig:dfa1dSist}
\end{figure}

With sinoartic denervation stationarity is lost, as DFA indicates
(Figure \ref{fig:dfa1dSist}). On pressure series from rats analyzed
$24$ hours after denervation (acute group) the crossover disappears,
and the series is non-stationary at all time scales ($\alpha \approx
1.25$), severely affecting homeostatic regulation of blood pressure.
Again we use higher order DFA check that no trends or
non-stationarities are shaping the results. The surrogate test is also
shown at the bottom curve of figure \ref{fig:dfa1dSist}a.

It is interesting to note that the same change of behavior has been
observed in the DFA analysis of fluctuations in blood glucose levels
of healthy humans and in patients with Diabetes Mellitus
\cite{AJP_RICP:291:R1638}: The damaged insulin metabolism controlling
blood sugar levels is reflected in the disappearance of the crossover
observed in the DFA curves of healthy subjects. In other study
\cite{ChaosSF:20:165}, this has been connected to the loss of
short-term adaptability of the cerebral blood flow control system of
migraineurs patients.

\begin{figure}[!htb]
\centering
\subfigure[]{
\includegraphics[scale=0.38]{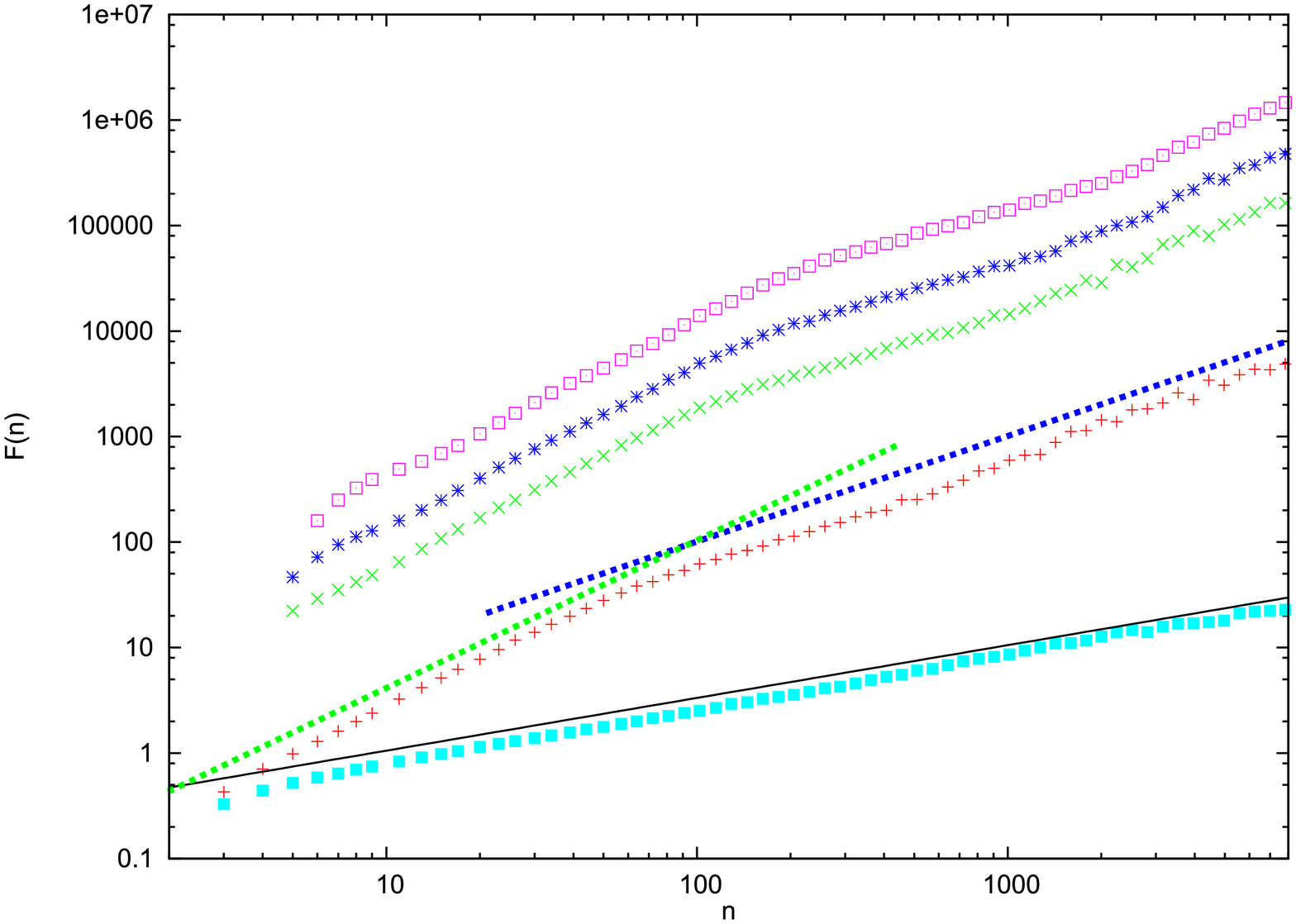}
\label{fig:dfa20dTip}
}
\subfigure[]{
\includegraphics[scale=0.45]{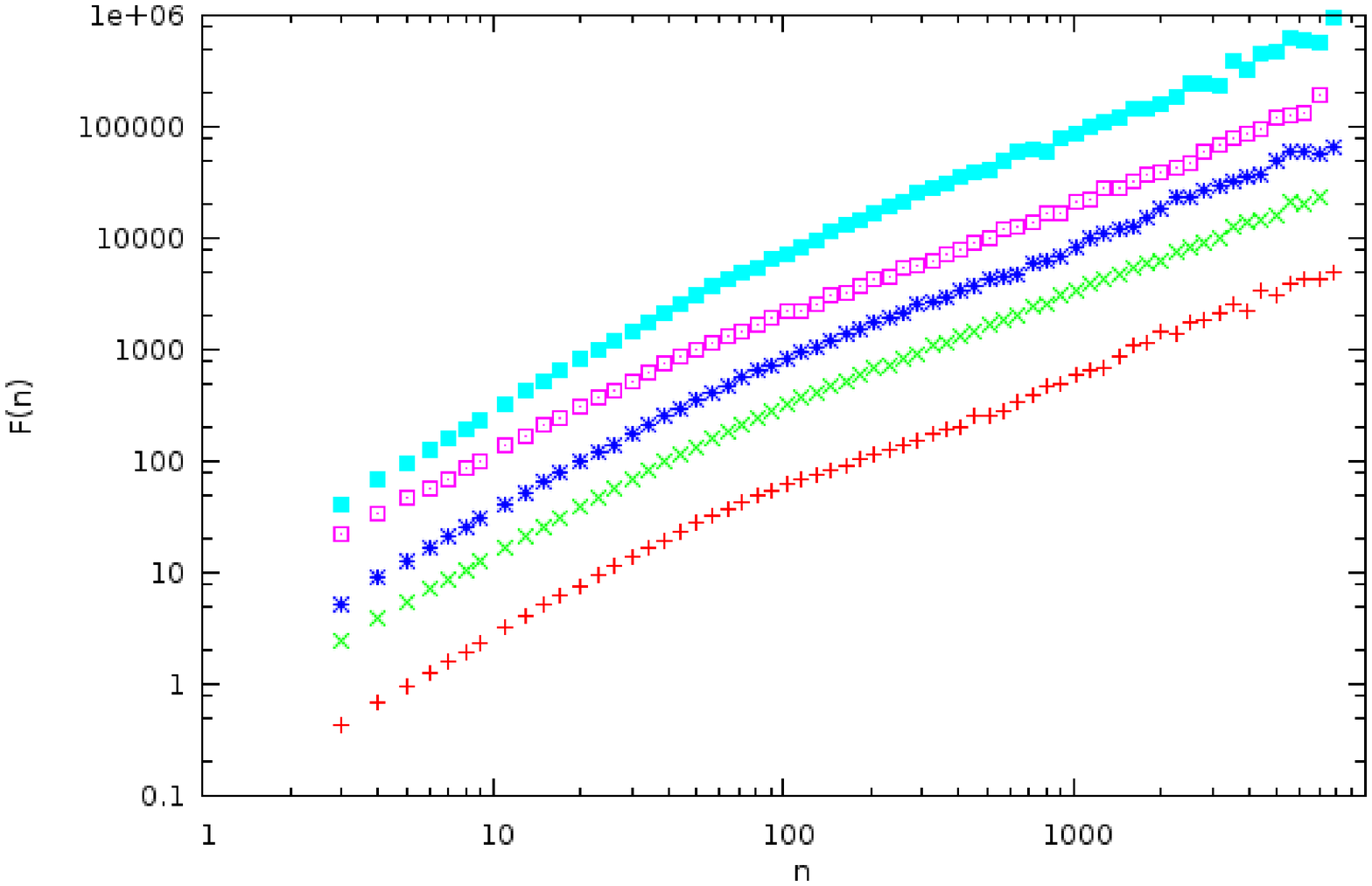}
\label{fig:dfa20dAll}
}
\caption{ {\bf\subref{fig:dfa20dTip}} Detrended fluctuation analysis
  of systolic blood pressure time series for a typical rat in the
  chronic group: $20$ days after surgical denervation, stationarity is
  recovered at large timescales and fluctuations cross over from
  non-stationary ($\alpha\approx 1.42$) to stationary, long-range
  correlated ($\alpha \approx 0.99$) at $n\approx 100$. This result
  suggests that, although the fast response from the baroreceptors in
  the heart is lost, physiological adaptation reestablishes
  homeostatic regulation.  We apply DFA-1 (red crosses), DFA-2 (green
  times), DFA-3 (blue stars) and DFA-4 (pink empty squares) to the
  series and find that the crossover always exists, although at
  different scales.  We also applied DFA-1 to shuffled data (bottom
  curve), for which $\alpha \approx 0.5$ as in white
  noise. {\bf\subref{fig:dfa20dAll}} DFA-1 for all rats in the chronic
  group. In both figures curves are shifted vertically for better
  visibility. The curves $y=Ax^{\alpha}$ with $\alpha=0.5$ (full black
  line), $1.0$ (dashed blue line) and $1.4$ (dashed green line) are
  plotted as guides to the eye.}
\label{fig:dfa20dSist}
\end{figure}

\begin{figure}[!hbt]
\centering
\includegraphics[scale=1.0]{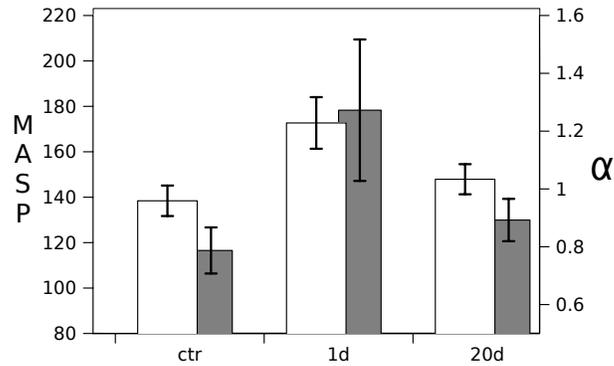}
\caption{Results for the mean arterial systolic pressure (MASP) (light
  gray bars) and the exponent $\alpha$ of long-term fluctuations
  averaged over all rats in each group, showing homeostasis adaptation
  of mean pressure and its fluctuations. In the control group (ctr),
  MASP have basal levels of $116.55 \pm 10.15$ mm Hg. One day after
  denervation (1d), MASP rises to $178.31 \pm 31.15$ mm Hg and, after
  20 days (20d), get back to a basal level of $129.95 \pm 9.32$,
  closer to basal levels of the control group. The long-range
  correlations observed both in control and chronic groups are
  statistically equivalent ($\alpha = 0.96 \pm 0.05$ for the first and
  $\alpha = 1.03 \pm 0.05$ for the latter group) with p-value
  $p=0.04$. Acute (1d) denervated rats have nonstationary
  fluctuations, with $\alpha = 1.23 \pm 0.09$ on all timescales.}
\label{fig:barrasHM}
\end{figure}

Twenty day past the denervation procedure, average blood pressure
returns to safe levels and stationarity is recovered (Figure
\ref{fig:dfa20dSist}): there is again a crossover from non-stationary
($\alpha \approx 1.42$) to stationary ($\alpha \approx 0.99$)
fluctuations, although at a larger timescale $n \approx 100$. Again we
use DFA-$1$ up to DFA-$4$ to insure that the crossover is not an
artifact of nonstationarities (Figure \ref{fig:dfa20dSist}a) and
depict in figure \ref{fig:dfa20dSist}b results for each rat in the
chronic group. The average blood pressure and the stationary,
long-range correlated fluctuations (as measured by $\alpha$ in the
region after the crossover) are statistically equivalent, as
summarized in figure \ref{fig:barrasHM}. When comparing the exponent
$\alpha$ in the control and chronic groups with a paired t-test
\cite{press} we find statistical equivalence with p-value $p=0.04$,
the same test for average blood pressure giving $p=0.07$.

\begin{figure}[!hbt]
 \centering
 \includegraphics[scale=0.3,angle=270]{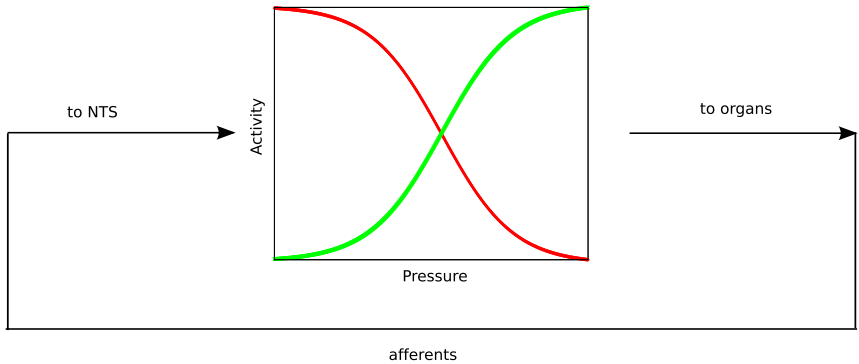}
 \caption{To model homeostatic blood pressure control we propose a
   simple model of a Brownian particle driven by noisy sigmoidal
   antagonistic forces $f_s$ (in red) and $f_v$ (in green).  Pressure
   information is sent through afferents to the NTS of the medulla,
   stimulating the vagal branch in a sigmoid fashion (green curve). The
   otherwise constant action of the sympathetic branch is modified by
   its vagal inhibition, resulting in the red curve depicted
   above. The equilibrium condition $f_s=f_v$ sets the average
   pressure.}
\label{fig:forces}
\end{figure}

Baroreflex recovery can be associated to the adaptation of sensory
neurons, most possibly at the Nucleus Tractus Solitarius (NTS)
\cite{AJP_RICP:289:R1416,JApplPhysio:101:322,RespPhysio:122:83} (the
mechanisms underlying this learning or synaptic plasticity are not
completely understood, but are already present in the adaptation of
stretch sensitivity in baroreceptors during the execution of simple
tasks such as sitting or head tilting for a reasonable amount of time
\cite{resetting_baro}). In rats with intact baroreceptors, baroreflex
sensitivity can be evaluated, both with vasoactive drugs or by
spontaneous fluctuations of heart rate and blood pressure, by means of
the \emph{Oxford method} \cite{CircRes:24:109}: Beat-to-beat variation
of systolic blood pressure is plotted against variation of the heart
rate at the subsequent beat interval. The slope of a linear regression
of this relation provides an index of arterial baroreflex sensitivity
(the same measure can be achieved by correlating blood flow and heart
rate variation and is known as the \emph{Trieste method}
\cite{trieste}). These methods assume that the two signals are
coupled, mostly at oscillatory frequencies of $0.4$ Hz
\cite{ANoninvElectr:13:191} and give a sigmoidal-like relation between
afferent nerve activity and blood pressure
\cite{AJP_HCP:291:H2003,resetting_baro}.  In order to model the action
of both vagal and sympathetic branches on blood pressure we devise a
model of a Brownian particle forced by opposing nonlinear forces, an
idea briefly touched upon in \cite{PhysRevE:70:050901}. Pressure
information merges through afferents and is integrated at the NTS,
stimulating the vagal branch, which further inhibits the sympathetic
branch of the autonomous nerve system. This coupled action can be
modelled by sigmoidal-like pressure-activity curves, as depicted in
figure \ref{fig:forces}: at each time step, pressure changes due to
the action of the forces $f_v(p)$ and $f_s(p)$ as
\begin{equation}
 p(t+1)=p(t)+\left(f_s(p+\xi(t))-f_v(p+\xi(t))\right)
\label{eq:model}
\end{equation}
where $\xi(t)$ represents the background noise integrated together
with afferent signal at the NTS, and the response curve $f_k(p)$ is
modeled by sigmoid-like curves
\cite{AJP_HCP:291:H2003,resetting_baro}:
\begin{equation}
f_k(p)=A_k \pm \frac{1}{B_k+e^{-(p-thr_{k})}}\label{eq:forces}
\end{equation}
where $k=s,v$ stands for sympathetic and vagal, respectively. In the
first case one subtracts and in the latter one adds the sigmoidal
curve to the base level of operation of each branch, called
\emph{tone}, represented by $A_k$. The parameters $thr_s$ and $thr_v$
give the pressure values for the optimal response of each branch: the
more different they are the larger is the region where pressure
fluctuates randomly.  In order to understand the role of the
antagonistic regulation of average blood pressure in our model, we
arbitrarily set $A_v=0.1$, $A_s=1.0$, $B_v=1.1$ and
$B_s=1.0$. \footnote{One could simplify the problem substituting the
  sigmoidal forces by step functions. We chose, however, to keep the
  biologically motivated sigmoidal responses.}

\begin{figure}[hbt]
  \centering
  \includegraphics[scale=0.4,angle = 270]{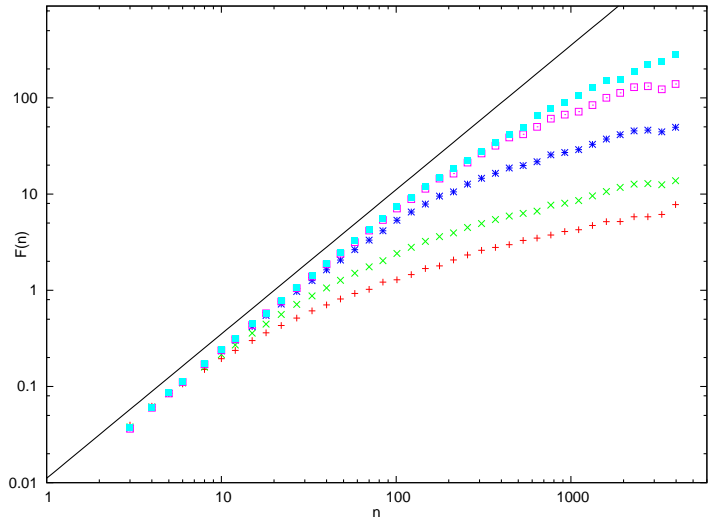}
  \caption{Detrended fluctuation analysis (DFA) of the artificial
    systolic blood pressure time series generated by the forced random
    walk model (equation \ref{eq:model}). We plot 5 values of the
    sensitivity difference: $thr_s - thr_v = 3$ (red cross),$thr_s -
    thr_v = 5$ (green times), $thr_s - thr_v = 8$ (blue stars),$thr_s
    - thr_v = 11$ (pink empty squares) and $thr_s - thr_v = 15$ (cyan
    full squares). The sensitivity difference increase as the same the
    crossover scale.  To guide the eye we show the curve with
    $\alpha=1.5$ (black full line). A large threshold for the action
    of autonomous system forces also means that more information
    (afferent signals) needs to be integrated at NTS to respond to a
    change in blood pressure.}
\label{fig:model}
\end{figure}

We analyze artificial systolic blood pressure series generated by such
forced random walk with DFA. After some transient behavior we store a
time series $\{p(t)\}$ with the same number of points as the
experimental datasets, $T=10^4$.  We find, with this simple model, the
same crossover observed in the actual pressure time series of intact
rats from the control group.  Moreover, keeping the same mechanism for
pressure control, but changing the sensitivity difference $thr_s -
thr_v$, we are able to reproduce the increase in the crossover scale
observed in chronic SAD rats (figure \ref{fig:model}).

This result can be understood by the following simple argument:
substituting the sigmoidal curves by step functions, the problem
reduces to one of a particle in a confining square-well potential of
width $L\approx thr_s - thr_v$. The first-passage-time of the random
walker to the walls of the potential sets a timescale for a crossover
between random, non-stationary fluctuations and confined motion
\cite{elsewhere}. Thus, with an increase of the width of the potential
well one should expect an increase of the range of the scaling region
related to non-stationary fluctuations. 

\section{Discussion}

We analyzed the dynamics of baroreflex, the negative feedback loop
providing reflex control of blood pressure by the autonomous nerve
system, with detrended fluctuation analysis of continuous
measurements of arterial systolic blood pressure. We report results of
our experiments with three groups of rats: a control group, another
group where baroreflex is surgically disrupted one day before
measurements and a third one, again with baroreflex surgically
impaired but whose measurements were made $20$ days after clinical
intervention. With DFA, we find on intact rats from the control group
a crossover from non-stationary to stationary, long-range correlated
fluctuations in arterial systolic blood pressure time series. This
crossover indicates that baroreflex sets in for pressure control at a
characteristic timescale.  One day after surgery one finds that the
feedback control, previously provided by baroreceptors, is impaired:
no crossover is found, and pressure fluctuations are
non-stationary. Nevertheless, after $20$ days of surgical intervention
we find evidence for physiological adaptation, and fluctuations scale
in a fashion which is statistically similar to those from the time
series of rats in the control group, only with the crossover from
non-stationary to stationary fluctuations occurring at a larger
timescale.  We also design a model for baroreflex which has the same
dynamical behavior of both normal and chronic SAD rats, qualitatively
reproducing the crossover in the scaling of fluctuations.  The main
feature of the model is its self-inhibitory behavior, which
illustrates the main principles underlying homeostatic control in
living systems, and has been observed at very different organizational
levels as an efficient mechanism for the maintenance of regularity in
a fluctuating environment.

\ack We thank A.Tavares Costa Jr. for a critical reading of the
manuscript and an anonymous referee for suggestions significantly
improving manuscript presentation.  This work is partially supported
by Brazilian Agencies CNPq, CAPES and FAPERJ.

\section*{References}
\bibliography{dfa} 

\end{document}